\documentclass[12pt]{article}
\usepackage{setspace}
\usepackage{amssymb}
\usepackage{amsmath,amssymb}
\usepackage{amsmath}
\usepackage[T1]{fontenc}
\usepackage[bf,small,tableposition=top]{caption}
\usepackage{subfig}
\usepackage{epsfig}
\usepackage{amsthm}
\usepackage{lineno}
\usepackage[figuresright]{rotating}
\usepackage{enumerate}
\usepackage{authblk}
\usepackage[T1]{fontenc}
\usepackage[utf8]{inputenc}
\usepackage{graphics}
\usepackage{graphicx}
\usepackage{multirow}
\usepackage{lscape}
\usepackage[dvipsnames]{xcolor}
\usepackage{mathrsfs}
\usepackage{epstopdf}
\usepackage{hyperref}
\usepackage{textcomp}
\usepackage{listings}
\usepackage{amsmath}
\usepackage{amsthm}
\usepackage{float}

\usepackage{tipa}
\usepackage{verbatim}
\usepackage{diagbox}
\usepackage{multirow,bigdelim,dcolumn,booktabs}

\theoremstyle{definition}

\theoremstyle{remark}

 \usepackage{multirow}
\begin{document}
\vspace{1cm}
\begin{center}
	\textbf{A Copula-Based family of Bivariate Composite Models for Claim Severity Modelling}
\end{center}

\begin{center}
	Girish Aradhye$^\dagger$, George Tzougas$^*$ and Deepesh Bhati$^\dagger$ \footnote{Corresponding author: deepesh.bhati@curaj.ac.in}\\

	$^\dagger$Department of Statistics, Central University of Rajasthan, Ajmer, India.\\
		$^*$Department of Actuarial Mathematics and Statistics, Heriot-Watt University, Edinburgh, EH14 4AS, United Kingdom. \\
\end{center}

\begin{abstract}

In this paper we consider bivariate composite models for modeling jointly different types of claims and their associated costs in a flexible manner. For expository purposes, the Gumbel copula is paired with the composite Weibull-Inverse Weibull, Paralogistic-Inverse Weibull and Inverse Burr-Inverse Weibull marginal models. The resulting bivariate copula-based composite models are fitted on motor insurance bodily injury and property damage data from a European motor insurance company and their parameters are estimated via the inference functions for margins method. \\ 

\end{abstract}

\section{Introduction}
Over the last few decades, there has been a vast increase of actuarial research works on modelling the cost of a specific claim type in non-life insurance based a variety of claim severity modelling approaches due to the peculiar characteristics of the claim severity distribution which poses several challenges as it often ranges over several magnitudes, from small and moderate claim sizes with a high frequency, as well as a few major ones with a low frequency. Additionally, claim size data are unimodal and heavily skewed to the right, see, for instance, Bakar et al. 2015). As it can be clearly understood, when the data spans over a wide range of magnitudes, selecting a probability distribution that can efficiently fit small and/or moderate and large claims becomes crucial for insurance pricing, reserving, and risk management.  Some popular methods that were developed in the literature for addressing this issue are  (i) the transformation of random variable (r.v.) (see, for example, Vernic, 2006, Adcock \textit{et al.}, 2015, Kazemi and Noorizadeh, 2015 and Eling, 2012, Bhati and Ravi, 2018), mixture of two or more distributions (Lin, 2010, Verbelen et al., 2015, Miljkovic and Gr\"{u}n, 2016), (ii) the method of compounding (see, for instance, Punzo et al., 2018) and (iii) the method of composition of distributions (see, for example, Cooray and Ananda, 2005,  Scollnik, 2007,  Ciumara,  2006, Cooray, 2009, Scollnik and Sun, 2012, Nadarajah and Bakar, 2014 and the references therein). Regarding the most recent studies on composite models, which are the main research focus of this work, it is worth noting that Nadarajah and Bakar (2014) proposed different composite models by considering the Burr, Loglogistic, Paralogistic and Generalized Pareto distributions for the tail of the data and truncated densities before and after the threshold point. Calderin-Ojeda and Kwok (2016) established alternative composite models by matching the two families of distributions at the modal value and they proposed the Lognormal-Stoppa and Weibull-Stoppa models for modelling the claim size data. Bhati et al. (2019) constructed composite models based on the Mode-Matching method and they considered different choices for the tail and head distribution. Grun and Miljkovic (2019), instead of creating composite models incrementally, consider 256 composite models that are derived from sixteen parametric distributions which are frequently used in actuarial science. Wang et al. (2020) put special emphasis on modelling extreme claims using a variety of composite models and threshold selection techniques, including heuristic methods, the Minimum AMSE of the Hill estimator, the exponentiality test, and the Gertensgarbe plot. Finally, Fung et al. (2022) introduce a mixture composite claim severity regression model extending the setup of Reynkens et al. (2017), who used a finite mixture distribution for the body  and a Pareto-type distribution for the tail of the distribution by incorporating explanatory variables on all three parts of the claim size distribution: clustering probabilities, body part, and tail part. At this point it should be noted that, even if the literature conserving composite models in the univariate setting is abundant, their bivariate extensions have not been investigated so far. Nevertheless, in non-life insurance it is common for the actuary to observe the existence of dependence structures between different types of claims and their associated costs either from the same type of coverage or from multiple types of coverage, such as, for example, motor and home insurance bundled into one single policy.
In this paper, motivated by a European Motor Third Party Liability (MTPL) insurance dataset, which is described in Section \ref{modelResult}, we introduce a family of bivariate composite models for joint modelling of the costs of positively correlated bodily injury and property damage claims. The proposed class of bivariate composite models is constructed by pairing a continuous copula distribution with two marginal composite distributions. The modelling framework we develop can account for the dependence structure between the two claim types in a versatile manner since it allows for a variety of alternative copula 
functions which can fully specify the dependence structure separately from the univariate marginal composite models (see, Joe 1997). Furthermore, depending on the choice of the composite marginal models, the proposed family  of bivariate composite models can be used to model the correlation between small and/or moderate and large claim sizes which can be the result of the same accident. 
We exemplify our approach by employing the Gumbel copula for modelling the dependence between bodily injury and property damage claims based on three bivariate composite models namely the  bivariate composite Weibull - Inverse Weibull, bivariate composite Paralogistic - Inverse Weibull and bivariate composite Inverse Burr - Inverse Weibull models. The parameters of the models are estimated using the inference function for margins (IFM) method whcih consists of estimating univariate parameters from separately maximizing the marginal composite models and then estimating the dependence parameters from the bivariate likelihoods which are derived based on the Gumbel copula.
  
\indent The rest of the paper is structured as follows. In section \ref{CC}, the alternative marginal composite models we consider herein are derived based on the  Classical Composition (CC) technique. Section \ref{bcm} presents the construction of the proposed bivariate composite models based on the Gumbel copula for modelling the dependence structure between different claim types. Model estimation via the IFM method along with the computational aspects of fitting the proposed bivariate composite models are discussed in section \ref{estimation}. In Section \ref{modelResult} we describe the MTPL dataset that we use for our empirical analysis and we provide estimation via the IFM method and model comparison for the proposed models. Finally, concluding remarks can be found in section \ref{conc}.\\

\section{Modelling framework : The Classical Composition Technique } \label{CC}
Bakar et al. (2015) proposed various composite models using unrestricted mixing weight ($r$), the right truncated and left truncated density truncated at threshold ($\theta$) for Head and Tail distributions respectively. The resulting probability density function (pdf) of the composite model can be written as
\begin{equation}
	f(x)=
	\begin{cases}
		rf^{*}_{1}(x|\Xi_1,\theta)&\text{for} \quad 0<x\le \theta,\\
		(1-r)f^{*}_{2}(x|\Xi_2,\theta)&\text{for} \quad \theta<x <\infty,\\
	\end{cases}
\end{equation}
where $r \in [0,1]$ and $\theta$ > 0. \\ The function $f^{*}_{1}(x|\Xi_1,\theta)=\frac{f_{1}(x|\Xi_1)}{F_{1}(\theta|\Xi_1)}$ and $f^{*}_{2}(x|\Xi_2,\theta)=\frac{f_{2}(x|\Xi_2)}{1-F_{2}(\theta|\Xi_2)}$
are the adequate truncation of the pdfs $f_{1}$ and 
$f_{2}$ upto and after an unknown threshold value $\theta$ respectively.
\begin{itemize}
	\item The value of weight parameter $r$ is obtained by continuity condition imposed at threshold $\theta$ i.e. $rf^{*}_{1}(\theta|\Xi_1,\theta)=(1-r)f^{*}_{2}(\theta|\Xi_2,\theta)$. Hence, we get\\\\
	\begin{equation} \label{rval}
		r=r(\theta,\Xi_1,\Xi_2)=\frac{f_{2}(\theta|\Xi_2)F_{1}(\theta|\Xi_1)}{f_{2}(\theta|\Xi_2)F_{1}(\theta|\Xi_1)+f_{1}(\theta|\Xi_1)(1-F_{2}(\theta|\Xi_2))}.
	\end{equation}
	
	\item Further, imposing the differentiability condition at threshold value $\theta$ $rf^{'}_{1}(\theta|\Xi_1,\theta)=(1-r)f^{'}_{2}(\theta|\Xi_2,\theta_2)$, makes the density smooth.  
\end{itemize}
These above conditions reduces the number of parameters and makes the resulting density continuous and differentiable. We henceforth refer this technique as Classical Composition (CC) technique.

\section{The Bivariate Composite Model} \label{bcm}
\indent Let $\mathbf{Y}=(\mathbf{Y^{(1)}},\mathbf{Y^{(2)}})$ and $\mathbf{y_i}=(y^{(1)}_i,y^{(2)}_i)$ respectively be the claims vector and its corresponding realizations of the two types of claims. Suppose that, $Y^{(j)}_i$, $j=1,2$ follows the composite $H$-Inverse Weibull (IW)  model with pdf given by
\begin{equation}\label{crmpdf}
	f_{j}(y^{(j)}_i)=
	\begin{cases}
			r^{(j)}_{H,IW}f_{H}^{*}(y^{(j)}_i;\Xi^{(j)}), & \text{for} \quad  0< y^{(j)}_i \le \theta^{(j)} \\\\
	(1-r^{(j)}_{H,IW}) \frac{\frac{\alpha^{(j)}}{y^{(j)}_i}\left(\frac{\gamma^{(j)}}{y^{(j)}_i}\right)^{\alpha^{(j)}}\exp\left\lbrace-\left(\frac{\gamma^{(j)}}{y^{(j)}_i}\right)^{\alpha^{(j)}}\right\rbrace}{1-\exp\left\lbrace-\left(\frac{\gamma^{(j)}}{\theta^{(j)}}\right)^{\alpha^{(j)}}\right\rbrace} &\text{for} \quad  \theta^{(j)} < y^{(j)}_i <\infty.
	\end{cases}
\end{equation} 

The cumulative distribution function (cdf) of composite $H$-Inverse Weibull   model may be written as
\begin{equation}\label{crmcdf}
	F_{j}(y^{(j)}_i)=
	\begin{cases}
		r^{(j)}_{H,IW}\frac{F_{H}(y^{(j)}_i;\Xi^{(j)}_1)}{F_{H}(\theta^{(j)};\Xi^{(j)}_1)}, &\text{for} \quad 0 < y^{(j)}_i \le \theta^{(j)}\\
		r^{(j)}_{H,IW}+(1-r^{(j)}_{H,IW}) \frac{\exp\left\lbrace-\left(\frac{\gamma^{(j)}}{y^{(j)}_i}\right)^{\alpha^{(j)}}\right\rbrace-\exp\left\lbrace-\left(\frac{\gamma^{(j)}}{\theta^{(j)}}\right)^{\alpha^{(j)}}\right\rbrace}{1-\exp\left\lbrace-\left(\frac{\gamma^{(j)}}{\theta^{(j)}}\right)^{\alpha^{(j)}}\right\rbrace} , & \text{for} \quad  \theta^{(j)} < y^{(j)}_i < \infty.
	\end{cases}
\end{equation}
 For $i=1,2 \cdots,n$ and $j=1,2$. Where $r^{(j)}_{H,IW} \in [0,1]$, $\Xi^{(j)} > 0$ are the parameters associated with the head density ($H$) of the marginal composite model, $\alpha^{(j)} > 0$, threshold point $\theta^{(j)} > 0$   and scale parameter $\gamma^{(j)} > 0$. $r^{(j)}_{H,IW}$ is the mixing weight associated with the right truncated $H$ density of the marginal composite $H$-Inverse Weibull model having $H$ distribution for the head part and Inverse Weibull distribution at the tail part. $F_H(y^{(j)})$ and $F_H(\theta^{(j)})$ are the cdf of the $H$ (Head) distribution evaluated at $y^{(j)}$ and the threshold point $\theta^{(j)}$ respectively. $f_H(.)$ is the density of  head part of the marginal composite $H$- Inverse Weibull model. $f_{H}^{*}(y^{(j)};\Xi^{(j)}_1)=\frac{f_{H}(y^{(j)};\Xi^{(j)}_1)}{F_{H}(\theta^{(j)};\Xi^{(j)}_1)}$ is the adequate right truncated density of the head part of the marginal composite $H$-Inverse Weibull model truncated at the threshold point $\theta^{(j)}$. 

We use following three distributions for modelling  the head part ($H$) of marginal composite $H$-Inverse Weibull models.
\begin{enumerate}
\item Weibull distribution:\\ $$f_{W}(y)=\frac{\mu}{\sigma}  \exp \left\lbrace -\left(\frac{y}{\sigma}\right)^{\mu}\right\rbrace \left(\frac{y}{\sigma }\right)^{\mu-1}, \quad \text{and} \quad F_{W}(y)=1-\exp \left\lbrace -\left(\frac{y}{\sigma}\right)^{\mu}\right\rbrace,$$\\
for $y > 0$, $\mu >0$, $\sigma > 0$ .

\item Paralogistic distribution:\\
	$$f_{P}(y)=\frac{\mu ^2 (y \sigma)^{\mu }}{y \left[(y\sigma)^{\mu }+1\right]^{\mu +1}}, \quad \text{and} \quad F_{P}(y)=\left[1-\left(\frac{1}{(\sigma y)^{\mu }+1}\right)^{\mu }\right].$$\\
	for $y > 0$, $\mu >0$ and $\sigma > 0$.

	\item Inverse Burr distribution: \\ $$f_{IB}(y)=\frac{\mu \sigma (y \tau)^{\mu  \sigma }}{y\left[(y \tau)^{\sigma }+1\right]^{\mu +1}}, \quad \text{and} \quad F_{IB}(y)=\left[\left((\tau  y)^{\sigma }+1\right)^{-\mu } (\tau  y)^{\mu  \sigma }\right],$$\\
	for $y > 0$, $\mu >0$, $\sigma > 0$ and $\tau > 0$.

\end{enumerate}
 By definition, a  copula  $C(u_1, u_2)$ is a bivariate cdf with uniform marginals on the interval $(0,1)$ (Joe 1997, Nelsen 2006). If $F_j(y_j)$ is the cdf of a univariate r.v. $Y_j$, then is $C(F_{1}(y^{(1)}_i), F_{2}(y^{(2)}_i))$ a bivariate distribution for $\mathbf{Y}=(\mathbf{Y^{(1)}},\mathbf{Y^{(2)}})$ with marginal distributions $F_j$, where $i=1,2\cdots,n$ and  $j=1,2$.\\
 The dependence of $Y_i$ among the claim types are modelled using copula, with the joint distribution of $Y_i$ given by
 \begin{equation}
     \pi(y_i)=C_{\Phi}\left(F_1\left(y^{(1)}_i\right),  F_2\left(y^{(2)}_i\right)\right),
 \end{equation}
where $\mathbf{y_i}=(y^{(1)}_i,y^{(2)}_i)$, $C_{\Phi}$ is a copula function and $\Phi := \{\phi^{(j,j')}\}_{j,j'=1,2}$ is the asymmetric parameter influencing the correlations among $(\mathbf{Y^{(1)}},\mathbf{Y^{(2)}})$.
The dependency between two types of claims $y^{(1)}_i,y^{(2)}_i$ can be modelled through the Gumbel copula as
\begin{equation}
    C_{\Phi}\left(F_1(y^{(1)}_i),  F_2(y^{(2)}_i)\right)=\exp \left\{-\left[\left(-\log F_1(y^{(1)}_i)\right)^{\phi}+\left(-\log F_2(y^{(2)}_i)\right)^{\phi}\right]^{\frac{1}{\phi}}\right\}.
\end{equation}

The detailed procedure involved in the generation of bivariate composite $H$-Inverse Weibull models for the three parametric distributions namely Weibull, Paralogistic and Inverse Burr distribution for the head distribution is given in the following section.

 \subsection{Bivariate Composite Weibull - Inverse Weibull (W-IW) Model}

Let $Y^{(j)}_i$ be the r.v. marginally follows a composite Weibull - Inverse Weibull model with pdf
\begin{equation}
	f_{j}(y^{(j)}_i)=
	\begin{cases}
			r^{(j)}_{W, IW}\frac{\frac{\mu^{(j)}}{\sigma^{(j)}}  \exp \left\lbrace -\left(\frac{y^{(j)}_i}{\sigma^{(j)}}\right)^{\mu^{(j)}}\right\rbrace \left(\frac{y^{(j)}_i}{\sigma^{(j)} }\right)^{\mu^{(j)}-1}}{1-\exp \left\lbrace -\left(\frac{\theta^{(j)}}{\sigma^{(j)}}\right)^{\mu^{(j)}}\right\rbrace}, & \text{for} \quad  0< y^{(j)}_i \le \theta^{(j)} \\\\
		(1-r^{(j)}_{W,IW}) (1-r^{(j)}_{H,IW}) \frac{\frac{\alpha^{(j)}}{y^{(j)}_i}\left(\frac{\gamma^{(j)}}{y^{(j)}_i}\right)^{\alpha^{(j)}}\exp\left\lbrace-\left(\frac{\gamma^{(j)}}{y^{(j)}_i}\right)^{\alpha^{(j)}}\right\rbrace}{1-\exp\left\lbrace-\left(\frac{\gamma^{(j)}}{\theta^{(j)}}\right)^{\alpha^{(j)}}\right\rbrace},   &\text{for} \quad \theta^{(j)} < y^{(j)}_i <\infty.
	\end{cases}
\end{equation} 
For $i=1,2,\cdots,n$ and $j=1,2$. $\mu^{(j)} > 0$, $\sigma^{(j)} > 0$, the scale parameter $\gamma^{(j)} > 0$, $\alpha^{(j)} > 0$, threshold point $\theta^{(j)} > 0$ , $r^{(j)}_{W, IW} \in [0,1]$ be the mixing weight of composite model which constitutes Weibull head and Inverse Weibull tail.
After imposing continuity condition at  threshold point $\theta^{(j)}$, we get 

\begin{equation}
    	r^{(j)}_{W, IW}=\frac{A}{A+B},
\end{equation}
where $A=\frac{\alpha^{(j)}}{\theta^{(j)}}\left(\frac{\gamma^{(j)}}{\theta^{(j)}}\right)^{\alpha^{(j)}}\exp\left\lbrace-(\frac{\gamma^{(j)}}{\theta^{(j)}})^{\alpha^{(j)}}\right\rbrace \left(1-\exp \left\lbrace -\left(\frac{\theta^{(j)}}{\sigma^{(j)}}\right)^{\tau^{(j)}}\right\rbrace\right)$ and\\ $B=\frac{\tau^{(j)}}{\sigma^{(j)}}  \exp \left\lbrace -\left(\frac{\theta^{(j)}}{\sigma^{(j)}}\right)^{\tau^{(j)}}\right\rbrace \left(\frac{\theta^{(j)}}{\sigma^{(j)} }\right)^{\tau^{(j)}-1} \exp\left\lbrace-\left(\frac{\gamma^{(j)}}{\theta^{(j)}}\right)^{\alpha^{(j)}}\right\rbrace$.\\\\

The cdf of composite W-IW   model can be written as
\begin{equation}\label{crmcdf}
	F_{j}(y^{(j)}_i)=
	\begin{cases}
		r^{(j)}_{W,IW}\frac{1-\exp \left\lbrace -\left(\frac{y^{(j)}_i}{\sigma^{(j)}}\right)^{\tau^{(j)}}\right\rbrace}{1-\exp \left\lbrace -\left(\frac{\theta^{(j)}}{\sigma^{(j)}}\right)^{\tau^{(j)}}\right\rbrace}, &\text{for} \quad 0 < y^{(j)}_i \le \theta^{(j)}\\
	r^{(j)}_{W,IW}+(1-r^{(j)}_{W,IW}) \frac{\exp\left\lbrace-\left(\frac{\gamma^{(j)}}{y^{(j)}_i}\right)^{\alpha^{(j)}}\right\rbrace-\exp\left\lbrace-\left(\frac{\gamma^{(j)}}{\theta^{(j)}}\right)^{\alpha^{(j)}}\right\rbrace}{1-\exp\left\lbrace-\left(\frac{\gamma^{(j)}}{\theta^{(j)}}\right)^{\alpha^{(j)}}\right\rbrace} , & \text{for} \quad  \theta^{(j)} < y^{(j)}_i < \infty.
	\end{cases}
\end{equation}
For $i=1,2,\cdots,n$ and $j=1,2$. The dependency among the two types of claims say $(y^{(1)}_i, y^{(2)}_i)$ can be studied using the Gumbel copula as follows
\begin{equation}
    C_{\phi}\left(F_1\left(y^{(1)}_i\right),  F_2\left(y^{(2)}_i\right)\right)=\exp \left\{-\left[\left(-\log F_1\left(y^{(1)}_i\right)\right)^{\phi}+\left(-\log F_2\left(y^{(2)}_i\right)\right)^{\phi}\right]^{\frac{1}{\phi}}\right\}.
\end{equation}
Where, $i=1,2,\cdots,n$ and $j=1,2$ and $y^{(1)}_i$ and $y^{(2)}_i$ are the realizations for the two types of claim $\mathbf{Y^{(1)}}$ and $\mathbf{Y^{(2)}}$. $F_1\left(y^{(1)}_i\right)$ and $F_2\left(y^{(2)}_i\right)$ are the cdfs of composite W-IW models evaluated at $y^{(1)}_i$ and $y^{(2)}_i$ respectively. $\phi$ is the asymmetric parameter control the correlations among $(\mathbf{Y^{(1)}}, \mathbf{Y^{(2)}})$.

 \subsection{Bivariate Composite Paralogistic-Inverse Weibull (P-IW) Model}

Let $Y^{(j)}_i$ be the r.v. marginally follows a composite Paralogistic-Inverse Weibull model with pdf
\begin{equation}
	f_{j}(y^{(j)}_i)=
	\begin{cases}
			r^{(j)}_{P, IW}\frac{\frac{(\mu^{(j)})^2 (y^{(j)}_i \sigma^{(j)})^{\mu^{(j)}}}{y^{(j)}_i \left[(y^{(j)}_i \sigma^{(j)})^{\mu^{(j)} }+1\right]^{\mu^{(j)} +1}}}{\left[1-\left(\frac{1}{(\sigma^{(j)} \theta^{(j)})^{\mu^{(j)}}+1}\right)^{\mu^{(j)}}\right]}, & \text{for} \quad  0< y^{(j)}_i \le \theta^{(j)} \\
		(1-r^{(j)}_{P,IW})  \frac{\frac{\alpha^{(j)}}{y^{(j)}_i}\left(\frac{\gamma^{(j)}}{y^{(j)}_i}\right)^{\alpha^{(j)}}\exp\left\lbrace-\left(\frac{\gamma^{(j)}}{y^{(j)}_i}\right)^{\alpha^{(j)}}\right\rbrace}{1-\exp\left\lbrace-\left(\frac{\gamma^{(j)}}{\theta^{(j)}}\right)^{\alpha^{(j)}}\right\rbrace},   &\text{for} \quad \theta^{(j)} < y^{(j)}_i <\infty.
	\end{cases}
\end{equation} 
For $i=1,2,\cdots,n$ and $j=1,2$. $\mu^{(j)} > 0$, $\sigma^{(j)} > 0$, the scale parameter $\gamma^{(j)} > 0$, $\alpha^{(j)} > 0$, threshold point $\theta^{(j)}$ > 0, $r^{(j)}_{P, IW} \in [0,1]$ be the mixing weight of composite model which constitutes Weibull head and Inverse Weibull tail.
After imposing continuity condition at  threshold point $\theta^{(j)}$, we get 

\begin{equation}
    	r^{(j)}_{W, IW}=\frac{A}{A+B},
\end{equation}
where, $A=\frac{\alpha^{(j)}}{\theta^{(j)}}\left(\frac{\gamma^{(j)}}{\theta^{(j)}}\right)^{\alpha^{(j)}}\exp\left\lbrace-(\frac{\gamma^{(j)}}{\theta^{(j)}})^{\alpha^{(j)}}\right\rbrace \left[1-\left(\frac{1}{(\sigma^{(j)} \theta^{(j)})^{\mu^{(j)} }+1}\right)^{\mu^{(j)} }\right]$ and \\

$B=\frac{(\mu^{(j)})^2 (\theta^{(j)} \sigma^{(j)})^{\mu^{(j)}}}{\theta^{(j)} \left[(\theta^{(j)} \sigma^{(j)})^{\mu^{(j)} }+1\right]^{\mu^{(j)} +1}} \exp\left\lbrace-\left(\frac{\gamma^{(j)}}{\theta^{(j)}}\right)^{\alpha^{(j)}}\right\rbrace$.\\

The cdf of composite P-IW  model can be written as
\begin{equation}\label{crmcdf}
	F_{j}(y^{(j)}_i)=
	\begin{cases}
		r^{(j)}_{P,IW}\frac{\left[1-\left(\frac{1}{(\sigma^{(j)} y^{(j)}_i)^{\mu^{(j)} }+1}\right)^{\mu^{(j)} }\right]}{\left[1-\left(\frac{1}{(\sigma^{(j)} \theta^{(j)})^{\mu^{(j)} }+1}\right)^{\mu^{(j)} }\right]}, &\text{for} \quad 0 < y^{(j)}_i \le \theta^{(j)}_j\\
	r^{(j)}_{P,IW}+(1-r^{(j)}_{P,IW}) \frac{\exp\left\lbrace-\left(\frac{\gamma^{(j)}}{y^{(j)}_i}\right)^{\alpha^{(j)}}\right\rbrace-\exp\left\lbrace-\left(\frac{\gamma^{(j)}}{\theta^{(j)}}\right)^{\alpha^{(j)}}\right\rbrace}{1-\exp\left\lbrace-\left(\frac{\gamma^{(j)}}{\theta^{(j)}}\right)^{\alpha^{(j)}}\right\rbrace} , & \text{for} \quad  \theta^{(j)} < y^{(j)}_i < \infty.
	\end{cases}
\end{equation} 
For $i=1,2,\cdots,n$ and $j=1,2$. The dependency among the two types of claims say $(y^{(1)}_i, y^{(2)}_i)$ can be studied using the Gumbel copula as follows
\begin{equation}
    C_{\phi}\left(F_1\left(y^{(1)}_i\right),  F_2\left(y^{(2)}_i\right)\right)=\exp \left\{-\left[\left(-\log F_1\left(y^{(1)}_i\right)\right)^{\phi}+\left(-\log F_2\left(y^{(2)}_i\right)\right)^{\phi}\right]^{\frac{1}{\phi}}\right\}.
\end{equation}
Where, $i=1,2,\cdots,n$ and $j=1,2$. $F_1\left(y^{(1)}_i\right)$ and $F_2\left(y^{(2)}_i\right)$ are the cdfs of composite P-IW models evaluated at $y^{(1)}_i$ and $y^{(2)}_i$ respectively.

 \subsection{Bivariate Composite Inverse Burr - Inverse Weibull (IB-IW) Model}

Let $Y^{(j)}_i$ be the r.v. marginally follows a composite Inverse Burr - Inverse Weibull model with pdf
\begin{equation}
	f_{j}(y^{(j)}_i)=
	\begin{cases}
			r^{(j)}_{IB, IW}\frac{\frac{\mu^{(j)} \sigma^{(j)} (y^{(j)}_i \tau^{(j)})^{\mu^{(j)}  \sigma^{(j)} }}{y^{(j)}_i \left[(y^{(j)}_i \tau^{(j)})^{\sigma^{(j)} }+1\right]^{\mu^{(j)} +1}}}{\left[\left((\tau^{(j)}  \theta^{(j)_j})^{\sigma^{(j)} }+1\right)^{-\mu^{(j)} } (\tau^{(j)}  \theta^{(j)})^{\mu^{(j)}  \sigma^{(j)} }\right]}, & \text{for} \quad  0< y^{(j)}_i \le \theta^{(j)} \\\\
		(1-r^{(j)}_{IB,IW}) \frac{\frac{\alpha^{(j)}}{y^{(j)}_i}\left(\frac{\gamma^{(j)}}{y^{(j)}_i}\right)^{\alpha^{(j)}}\exp\left\lbrace-\left(\frac{\gamma^{(j)}}{y^{(j)}_i}\right)^{\alpha^{(j)}}\right\rbrace}{1-\exp\left\lbrace-\left(\frac{\gamma^{(j)}}{\theta^{(j)}}\right)^{\alpha^{(j)}}\right\rbrace},   &\text{for} \quad \theta^{(j)} < y^{(j)}_i <\infty.
	\end{cases}
\end{equation} 
For $i=1,2,\cdots,n$ and $j=1,2$. $\mu^{(j)} > 0$, $\sigma^{(j)} > 0$, $\tau^{(j)} > 0$, the scale parameter $\gamma^{(j)} > 0$, $\alpha^{(j)} > 0$, threshold point $\theta^{(j)}$ > 0, $r^{(j)}_{IB, IW} \in [0,1]$ be the mixing weight of composite model which constitutes Inverse Burr head and Inverse Weibull tail.

The analytical expression for the mixing weight $r^{(j)}_{IB, IW}$ can be written as 

\begin{equation}
    	r^{(j)}_{W, IW}=\frac{A}{A+B},
\end{equation}
where $A=\frac{\alpha^{(j)}}{\theta^{(j)}}\left(\frac{\gamma^{(j)}}{\theta^{(j)}}\right)^{\alpha^{(j)}}\exp\left\lbrace-(\frac{\gamma^{(j)}}{\theta^{(j)}})^{\alpha^{(j)}}\right\rbrace \left[\left((\tau^{(j)}  \theta^{(j)})^{\sigma^{(j)} }+1\right)^{-\mu^{(j)} } (\tau^{(j)}  \theta^{(j)})^{\mu^{(j)}  \sigma^{(j)} }\right]$ and \\
$B=\frac{\mu^{(j)} \sigma^{(j)} (\theta^{(j)} \tau^{(j)})^{\mu^{(j)}  \sigma^{(j)} }}{\theta^{(j)} \left[(\theta^{(j)} \tau^{(j)})^{\sigma^{(j)} }+1\right]^{\mu^{(j)} +1}} \exp\left\lbrace-\left(\frac{\gamma^{(j)}}{\theta^{(j)}}\right)^{\alpha^{(j)}}\right\rbrace$.\\\\

The cdf of composite IB-IW   model can be written as
\begin{equation}\label{crmcdf}
	F_{j}(y^{(j)}_i)=
	\begin{cases}
		r^{(j)}_{P,IW}\frac{\left[\left((\tau^{(j)}  y^{(j)}_i)^{\sigma^{(j)} }+1\right)^{-\mu^{(j)} } (\tau^{(j)}  y^{(j)}_i)^{\mu^{(j)}  \sigma^{(j)} }\right]}{\left[\left((\tau^{(j)}  \theta^{(j)})^{\sigma^{(j)} }+1\right)^{-\mu^{(j)} } (\tau^{(j)}  \theta^{(j)})^{\mu^{(j)}  \sigma^{(j)} }\right]}, &\text{for} \quad 0 < y^{(j)}_i \le \theta^{(j)}\\
	r^{(j)}_{IB,IW}+(1-r^{(j)}_{IB,IW}) \frac{\exp\left\lbrace-\left(\frac{\gamma^{(j)}}{y^{(j)}_i}\right)^{\alpha^{(j)}}\right\rbrace-\exp\left\lbrace-\left(\frac{\gamma^{(j)}}{\theta^{(j)}}\right)^{\alpha^{(j)}}\right\rbrace}{1-\exp\left\lbrace-\left(\frac{\gamma^{(j)}}{\theta^{(j)}}\right)^{\alpha^{(j)}}\right\rbrace}, & \text{for} \quad  \theta^{(j)} < y^{(j)}_i < \infty.
	\end{cases}
\end{equation} 
 For $i=1,2,\cdots,n$ and $j=1,2$. The dependency among the two types of claims say $(y^{(1)}_i,y^{(2)}_i)$ can be studied using the Gumbel copula as follows
\begin{equation}
    C_{\phi}\left(F_1\left(y^{(1)}_i\right),  F_2\left(y^{(2)}_i\right)\right)=\exp \left\{-\left[\left(-\log F_1\left(y^{(1)}_i\right)\right)^{\phi}+\left(-\log F_2\left(y^{(2)}_i\right)\right)^{\phi}\right]^{\frac{1}{\phi}}\right\}.
\end{equation}
Where, $i=1,2,\cdots,n$ and $j=1,2$. $F_1\left(y^{(1)}_i\right)$ and $F_2\left(y^{(2)}_i\right)$ are the cdfs of composite IB-IW models evaluated at $y^{(1)}_i$ and $y^{(2)}_i$ respectively.

\section{Parameter Estimation via the IFM Method} \label{estimation}
The objective of this section is to go over how to use the maximum likelihood (ML) method to estimate the parameters related to marginals as well as copula. Joe (1997) discussed a method known as inference function for margins (IFM) for parameter estimation of the copula density, which is dependent on the knowledge of the marginals. This method has two steps in which the parameters of the marginal df's are estimated first and then the copula parameters are obtained by maximising the likelihood function of the copula with the marginal parameters replaced by estimators obtained in the first step. The steps involved in the inference function for margins (IFM) method for the estimation of marginal model parameters as well as the copula parameter are given below.
\begin{itemize}
    \item \textbf{Step 1:}
Let $Y^{(j)}_{1}, Y^{(j)}_{2},\cdots, Y^{(j)}_{n}$ be a random sample of two types of claims form the  marginal composite $H$-Inverse Weibull  model described in  \ref{crmpdf} for $j=1,2$. Here $\mathbf{\Theta}$ is the parameter vector for the two marginal composite $H$-Inverse Weibull  models. We utilize the maximum likelihood (ML) estimation procedure to estimate the marginal model parameters. The goal of maximum likelihood estimation procedure is to find the values for parameters which maximize 
\begin{equation*}
	l(\mathbf{\Theta^{(j)}}|y^{(j)}_i)= \sum_{i=1}^{n}\text{ln}(f_{j}(y^{(j)}_{i}|\mathbf{\Theta}))
\end{equation*}
\begin{align*} \nonumber
		& l(\mathbf{\Theta^{(j)}}|y^{(j)}_{i})= \sum_{i=1}^{n}\text{ln} \left[		r^{(j)}_{H,IW}\frac{f_{H}(y^{(j)}_{i};\Xi^{(j)})}{F_{H}(\theta;\Xi^{(j)})} \mathbb{I}[y_{i} < \theta^{(j)}] \right.\\
	&  \left. \qquad +  	(1-r^{(j)}_{H,IW}) \frac{\alpha^{(j)}}{y^{(j)}_{i}}\left(\frac{\gamma^{(j)}}{y^{(j)}_{i}}\right)^{\alpha^{(j)}}\exp\left\lbrace-\left(\frac{\gamma^{(j)}}{y^{(j)}_{i}}\right)^{\alpha^{(j)}}\right\rbrace \mathbb{I}[y^{(j)}_{i} \ge \theta^{(j)})]\right] 
\end{align*}
\begin{align} \nonumber
		 l(\mathbf{\Theta^{(j)}}|y^{(j)}_i)= &\sum_{i=1}^{n} \left[\ln r^{(j)}_{H,IW}\mathbb{I}[y^{(j)}_{i} < \theta^{(j)}]+\ln f_{H}(y^{(j)}_{i};\Xi^{(j)})\mathbb{I}[y^{(j)}_{i} < \theta^{(j)}]\right. \\ \nonumber 
		 & \qquad \left.-\ln F_{H}(\theta^{(j)};\Xi^{(j)})\mathbb{I}[y^{(j)}_{i} < \theta^{(j)}]  +  \ln	(1-r^{(j)}_{H,IW}) \mathbb{I}[y^{(j)}_{i} \ge \theta^{(j)}] \right. \\
			&  \qquad 
	\left.+\ln \alpha^{(j)} -\ln  y^{(j)}_{i}+\alpha^{(j)}(\ln \gamma^{(j)}-\ln y^{(j)}_{i})-\left(\frac{\gamma^{(j)}}{y^{(j)}_{i}}\right)^{\alpha^{(j)}}\mathbb{I}[y^{(j)}_{i} \ge \theta^{(j)}] \right].
\end{align}

For $j=1,2$. All of the parameters of  composite $H$-Inverse Weibull model are estimated using the numerical optimization tool \texttt{optim()}, included in the \texttt{stats} package of \texttt{R} programming language.

\item \textbf{Step 2:} The estimated parameters of the marginal df's are used to estimate the Gumbel copula parameter and to compute the value of likelihood function associated with the dependence structure. To ease the process of finding the estimate of copula parameter we use the \texttt{rvinecopula} package of \texttt{R} software by passing the argument  \texttt{gumbel} as a   \texttt{family}.
\end{itemize}

\section{Motivating Dataset: Greek MTPL Bodily Injury and Property Damage Claim Amounts} \label{modelResult}
\begin{figure}[H]
	\centering
\includegraphics[width=9.5cm]{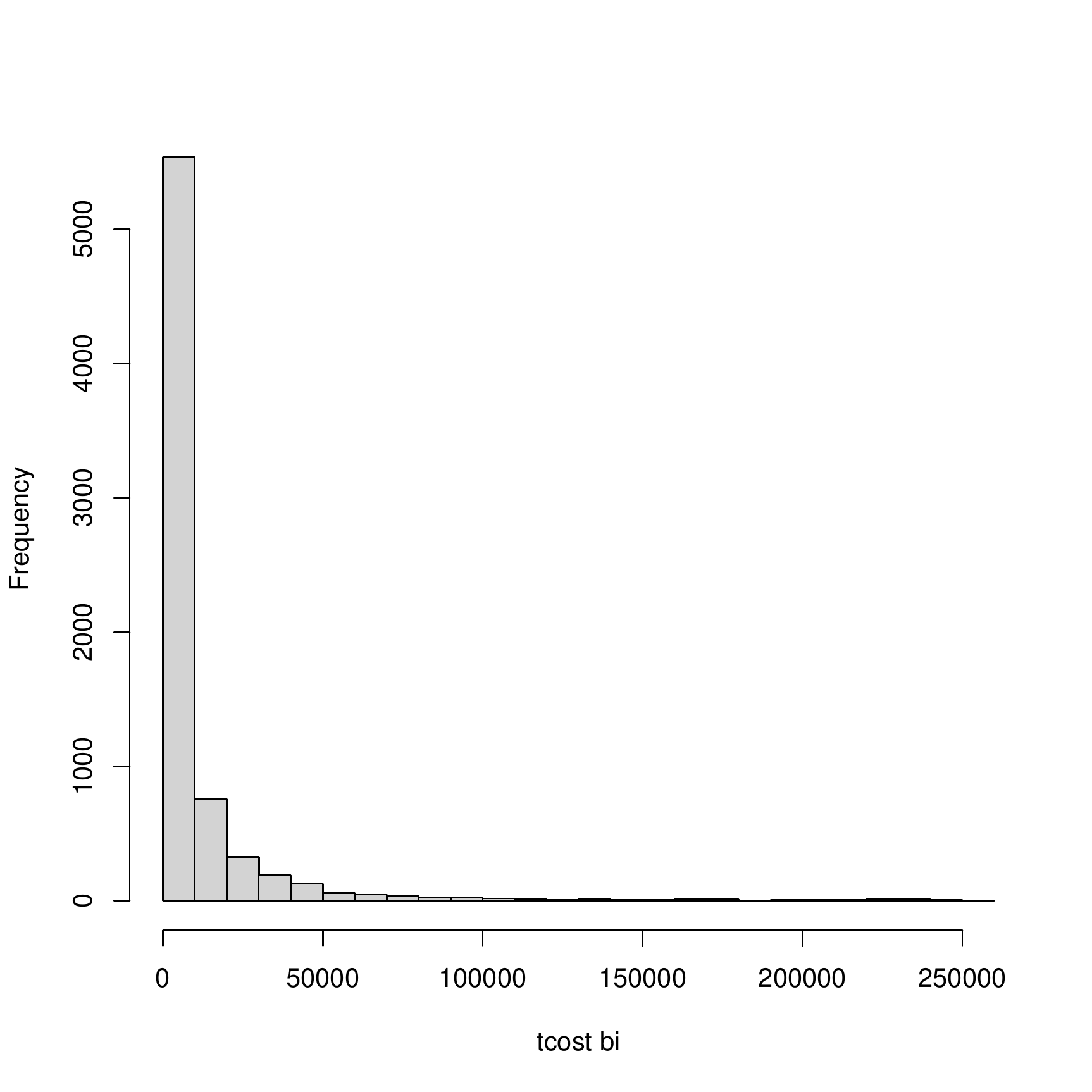}
\caption{Histogram of the bodily injury claims for the \texttt{MTPL} dataset}
	\label{histbiclaim}
\end{figure}

\begin{figure}[H]
	\centering
\includegraphics[width=9.5cm]{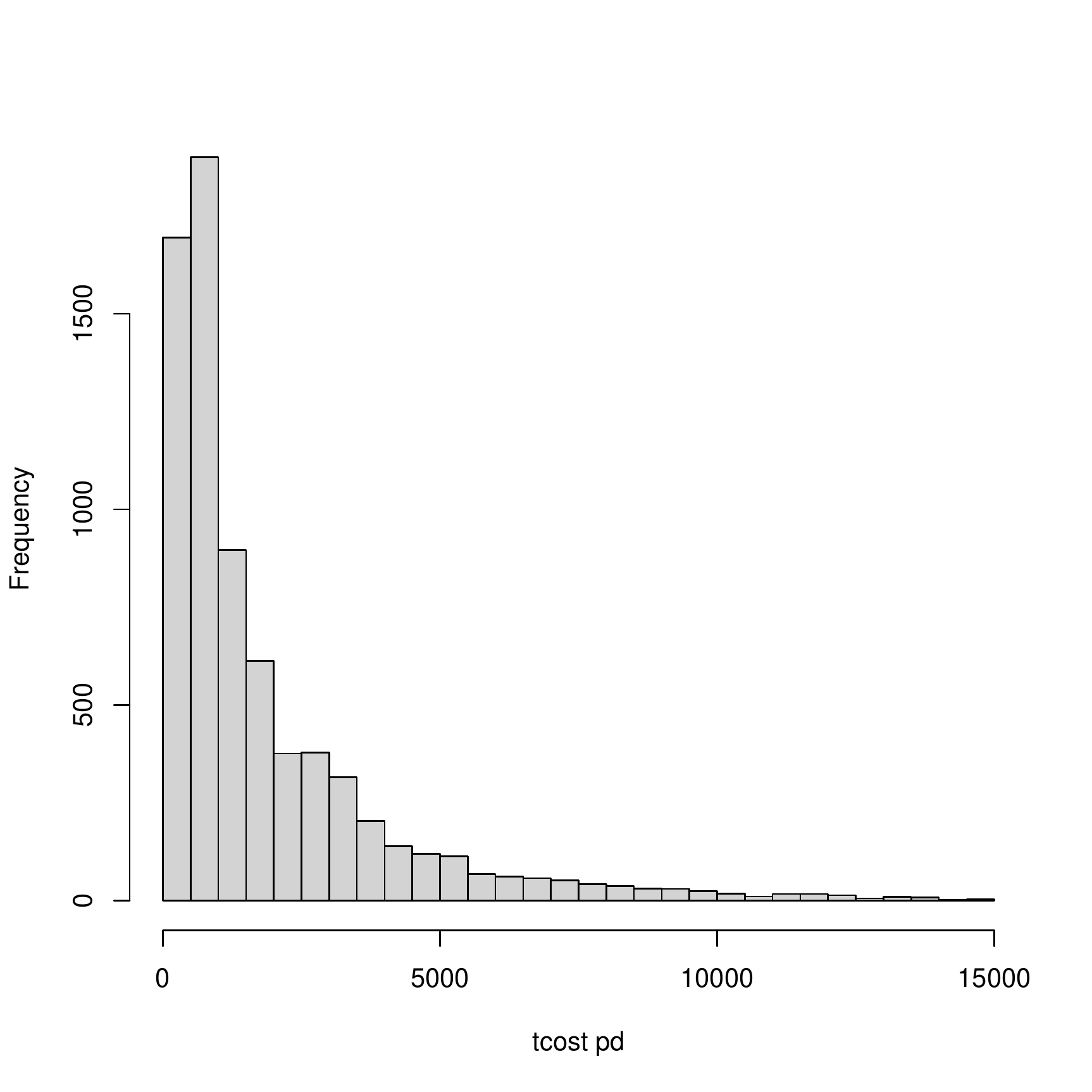}
\caption{Histogram of the property damage claims for the \texttt{MTPL} dataset}
	\label{histpdclaim}
\end{figure}

\begin{figure}[H]
	\centering
\includegraphics[width=9.5cm]{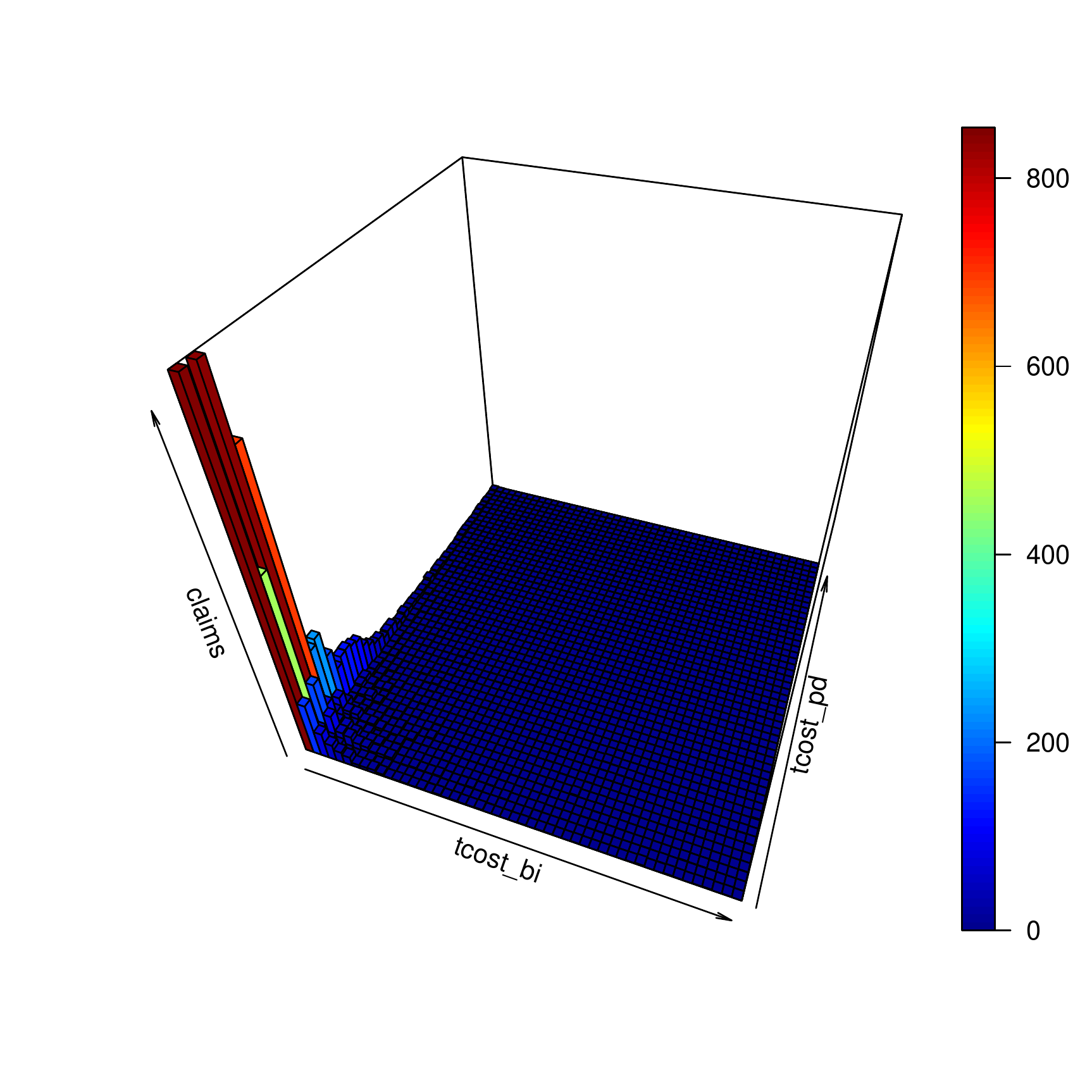}
\caption{Histogram of the bodily injury claims and property damage claims for the \texttt{MTPL} dataset}
	\label{histbipdclaim}
\end{figure}

In this section, we illustrate the proposed methodology using the motor third-party liability (\texttt{MTPL}) insurance policies with non-zero property claims for the years 2012 to 2019. A major insurance firm in Greece generously provided the dataset for this study. The dataset contains 7263 policies of motor vehicle insurance collected over a period in 2012-2019 which have complete records. Following section provides the detailed description of the (\texttt{MTPL}) dataset. The  variables associated with the \texttt{MTPL} dataset is \texttt{tcost bi} i.e. the cost of bodily injury claims and \texttt{tcost pd} i.e. the cost of property damage  claims are a numeric vectors showing the total amount of bodily injury claims and property damage claims. Table \ref{srce} depicts the descriptive statistics for \texttt{tcost bi}  and \texttt{tcost pd}. Graphical representations of the both types of claims namely bodily injury claims and property damage claims are presented in Figure  \ref{histbiclaim}, Figure \ref{histpdclaim} and Figure \ref{histbipdclaim}. Both the types of claims exhibits the various peculiarities of the insurance data including positive skewness, unimodal, tail heaviness. 

  \begin{table}[H]
	\centering
	\caption{Summary of \texttt{tcost bi} and \texttt{tcost pd} claim amounts of the \texttt{MTPL} dataset.}
	\scalebox{1.01}{
	\begin{tabular}{ccccccccc} \hline \hline
		 Variable & Minimum & Maximum & \texttt{Q1} & Median & \texttt{Q3}& Mean & Skewness & Kurtosis \\ \hline \hline
	  \texttt{tcost bi}&  0.9 & 251958.2 & 318.0 & 2413.4 & 9120.6 & 11017.3 & 5.31 & 36.88     \\
	  \texttt{tcost pd} & 6.2 & 14818.2  & 584.3 & 1012.4  & 2461.8 & 1871.2 & 2.38 & 9.71  \\
		 \hline
	\end{tabular}}
	\label{srce}
\end{table}

\subsection{Fitting Results and Model Comparison}
In this section, we present the results based on the two model selection criterion for the proposed models. Table \ref{NLL}  provide the values of Akaike’s information criterion (AIC) (see Akaike, 1974) and Bayesian information criterion (BIC). Note that for AIC and BIC, smaller values indicate a better fit of the model to the empirical data. The formulae involved in the computation of above mentioned model selection criterion's are\\
The AIC can be computed as
 \begin{equation*}
 	AIC=-2{l}(\hat{\mathbf{\Theta}})+ 2 \times df,
 \end{equation*} 
 where ${l(\hat{\mathbf{\Theta}})}$ is the maximum of the log-likelihood and $\hat{\mathbf{\Theta}}$ is the vector of the estimated model parameters. 
and the BIC is given by
 \begin{equation*}
	BIC=-2{l}(\hat{\mathbf{\Theta}})+ \text{log}(n) \times df,
\end{equation*}
where $n$ is sample size of the dataset and $df$ is the number of fitted parameters of the model.\\
The following results from Table \ref{NLL} shows that the Bivariate composite Paralogistic -Inverse Weibull model performs better than remaining  Bivariate composite  models with  followed by Bivariate composite Inverse Burr-Inverse Weibull  model. Table \ref{paraest} presents the parameter estimates (marginal model parameters and copula parameter) of the proposed composite models for the \texttt{MTPL} dataset.

\begin{table}[H]
	\centering
	\caption{Values of AIC and BIC for \texttt{MTPL} dataset }
\begin{tabular}{ccccc} \hline \hline
\multicolumn{2}{c}{Model}      &       &      &      \\
Head         & Tail            & Parameters & AIC       & BIC       \\ \hline \hline
Paralogistic & Inverse Weibull & 9          & 265365.91 & 265428.01 \\
Inverse Burr & Inverse Weibull & 11          & 266858.41 & 266934.21 \\
Weibull      & Inverse Weibull & 9         & 270218.81 & 270280.81 \\  \hline \hline
\end{tabular}
	\label{NLL}
\end{table}

\begin{table}[H]
\centering
\caption{Parameter estimates of composite models for \texttt{tcost bi} claims of \texttt{MTPL} dataset }
 \scalebox{0.8}{
\begin{tabular}{ll|lllllll}  \hline  \hline
\multicolumn{2}{c}{Model}      & \multicolumn{7}{c}{tcost bi}                                        \\  \hline  \hline
Head         & Tail            &    $\mu^{(1)}$     & $\sigma^{(1)}$   & $\tau^{(1)}$    & $\alpha^{(1)}$  & $\gamma^{(1)}$     & $r^{(1)}$      & $\theta^{(1)}$    \\  \hline  \hline
Paralogistic & Inverse Weibull & 0.7991 & 0.0008  & --     & 0.9046 & 10034.72  & 0.8832 & 25449.23  \\
Inverse Burr & Inverse Weibull & 0.0002 & 1658.49 & 0.0003 & 1.0264 & 4281.5141 & 0.5658 & 3877.8737 \\
Weibull      & Inverse Weibull & 0.5394 & 4644.45 & --    & 1.3988 & 13751.62  & 0.9102 & 27179.21 \\  \hline  \hline
\end{tabular}}
\end{table}

\begin{table}[H]
  \centering
  \caption{Parameter estimates of composite models for \texttt{tcost pd} claims and Copula Parameter of \texttt{MTPL} dataset }
    \scalebox{0.8}{
\begin{tabular}{ll|llllllll} \hline \hline
\multicolumn{1}{c}{Model} & \multicolumn{1}{c}{} & \multicolumn{7}{c}{tcost pd}                                                                                     & $\phi$ \\ \hline \hline
Head                      & Tail                 & $\mu^{(2)}$     & $\sigma^{(2)}$                                                & $\tau^{(2)}$    & $\alpha^{(2)}$  & $\gamma^{(2)}$      & $r^{(2)}$      & $\theta^{(2)}$    &                  \\ \hline \hline
Paralogistic              & Inverse Weibull      & 1.2596 & 0.0007                                               & --    & 2.4474 & 11634.1078 & 0.9865 & 16941.38 & 0.1196           \\
Inverse Burr              & Inverse Weibull      & 0.0301 & 34.9757                                              & 0.0018 & 0.8872 & 430.1964   & 0.3531 & 499.9917 & 0.1495           \\
Weibull                   & Inverse Weibull      & 0.5485 & 1.2 $\times 10^{11}$ & --    & 0.5139 & 410.9973   & 0.2191 & 100.0001 & 0.2791  \\ \hline        
\end{tabular}}
\end{table}

\subsection{Analysis of Dependence}
To examine the goodness-of-fit of the fitted model in terms of dependence modelling, we provided the values of Kendall's tau between empirical data and the fitted models under Gumbel copula. From Table \ref{ktau} it is clear that, in terms of Kendall's taus, the fitted model matches the actual data reasonably well, implying that the suggested copula model can properly reflect the dependence structure of the both type claims of the \texttt{MTPL} dataset.
\begin{table}[H]
	\centering
	\caption{Kendall's tau for the fitted model versus the empirical dataset}
\begin{tabular}{cccc}\hline \hline
\multicolumn{2}{c}{Model}      &           & \multicolumn{1}{l}{}              \\
Head         & Tail            & Kendall's tau & \multicolumn{1}{l}{Empirical dataset} \\\hline \hline
Paralogistic & Inverse Weibull & 0.0663        & \multirow{3}{*}{0.0997}               \\
Inverse Burr & Inverse Weibull & 0.0159        &                                       \\
Weibull      & Inverse Weibull & 0.1319       &    \\\hline \hline                                  
\end{tabular}
\label{ktau}
\end{table}

\section{Conclusions} \label{conc}
In this paper we presented a novel family of bivariate composite models for simultaneously modelling small and/or moderate and large costs from different types of claims in an efficient manner based on the choice of alternative composite marginal densities and a copula model for specifying the correlation structure between the claim size response variables. We illustrated our approach by fitting on MTPL bodily injury and property damage data the  bivariate composite models which were derived by pairing the Weibull-Inverse Weibull, Paralogistic-Inverse Weibull and Inverse Burr-Inverse Weibull models using the Gumbel copula which also enabled us to investigate the right tail dependence. The model parameters as well as the dependence parameter were estimated based on the inference function for margins (IFM) method. 
A potentially fruitful line of further research is to  consider the influence of individual and coverage specific covariates on the mean, dispersion and dependence parameters of the porposed class of models by introducing regression components into the composite marginal models and the correlation parameter of the Gumbel copula.

\section*{Funding}
The present work is a part of project granted by Department of Science \& Technology, Government of India under the Core Research Grant scheme (CRG/2019/002993). D.B. and G.A. thank the funding agency for financial support.

\end{document}